\documentclass[a4paper,fleqn,10pt]{article}

\newtheorem{thm}{Theorem}[section]

\newtheorem{lem}[thm]{Lemma}

\newtheorem{dfn}[thm]{Definition}
\newtheorem{hyp}[thm]{Hypotheses}

\usepackage[T1]{fontenc}
\usepackage[english]{babel}
\usepackage{amsfonts}
\usepackage{amsmath}
\usepackage[left=20mm, right=20mm, top=25mm, bottom=25mm]{geometry}
\usepackage[pdftex]{hyperref}
\usepackage{cite}

\title{The singular kernel coagulation equation with multifragmentation}
\author{Carlos Cueto Camejo\thanks{Corresponding author. Fax: +49 391 6718073,\newline  E-mail address: karlos.cueto@gmail.com (C. Cueto Camejo).}, Gerald Warnecke \\ \small{\emph{Institute for Analysis and Numerics, Otto-von-Guericke University Magdeburg,}} \\ \small{\emph{Universit\"atsplatz 2, D-39106 Magdeburg, Germany}}}

\begin{document}
\maketitle

\begin{abstract}
In this article we prove the existence of solutions to the singular coagulation equation with 
multifragmentation. We use weighted $L^1$-spaces to deal with the singularities and to obtain regular solutions. The 
Smoluchowski kernel is covered by our proof. The  weak $L^1$ compactness methods are applied 
to suitably chosen approximating equations as a base of our proof. A more restrictive 
uniqueness result is also given.
\end{abstract}

\section{Introduction}
The coagulation process describes the kinetics of particle growth where particles can coagulate to form larger particles via binary interaction. On the other side, the fragmentation process describes how particles break into two or more fragments. Examples of these processes can be found e.g. in astrophysics, polymer science \cite{ziff}, and cloud physics \cite[Chapter 15]{Pruppacher}.

The dynamic of the coagulation-fragmentation process is described by the integro-differential equation 
\begin{eqnarray}
&&\frac{\partial u(x,t)}{\partial t}=\frac{1}{2}\int\limits_0^x K(x-y,y)u(x-y,t)u(y,t)\,dy-\int\limits_0^\infty K(x,y)u(x,t)u(y,t)\,dy \nonumber\\
&&\qquad\quad\qquad+\int\limits_x^\infty b(x,y)S(y)u(y,t)\,dy-S(x)u(x,t),
\label{Fproblema}
\end{eqnarray}
with initial condition
\begin{eqnarray}
u(x,0)=u_0(x)\geq 0\quad \mbox{a.e.}
\label{Fcondinicial}
\end{eqnarray}
where the non-negative variables $x$ and $t$ represent the size of the particles and time respectively. The values $u(x,t)$ denote the number density of particles with size $x$ at time $t$. The rate at which particles of size $x$ coalesce with particles of size $y$ is represented by the coagulation kernel $K(x,y)$. The rate at which particules of size $x$ are selected to break is determined by the selection function $S(x)$. The breakage function $b(x,y)$ gives the number of particules of size $x$ produced when a particule of size $y$ breaks up.

The equation (\ref{Fproblema}) is named the continuous coagulation equation with multifragmentation where the multifragmentation kernel $\Gamma$ defines the selection function $S$ and the breakage function $b$ by
\begin{eqnarray}
S(x)=\int\limits_0^x\frac{y}{x}\Gamma(x,y)dy, \qquad\quad b(x,y)=\Gamma(y,x)/S(y),
\label{FB}
\end{eqnarray}
or vice versa.

The breakage function is assumed here to have the following properties
\begin{eqnarray}
\int\limits_0^yxb(x,y)dx=y\quad\mbox{for all}\quad y>0,
\label{Fb}
\end{eqnarray}
which is the conservation of mass and
\begin{eqnarray}
\int\limits_0^yb(x,y)dx=N<\infty\quad\mbox{for all}\quad y>0,\; b(x,y)=0\quad\mbox{for}\quad x>y,
\label{FbN}
\end{eqnarray}
where the parameter $N$ represents the number of particles produced in fragmentation events. In 
this paper $N$ is assumed to be finite and independent of $y$. Equation (\ref{Fb}) allows the 
system to conserve the total mass during the fragmentation events. It states that the total mass 
of the fragments  is equal to the mass $y$ of the particle that breaks.

The existence and uniqueness of solutions to the coagulation-fragmentation equation has already 
been the subject of several papers. The case of multifragmentation, that is, when the particules 
can break into two or more parts, has also been studied, see e.g.\ \cite{PhLaurencot}, \cite{McLaughlin}, 
\cite{Melzak}, and \cite{Walker}.
For more recent result see e.g.\ \cite{BanasiakLamb}, \cite{LambBanasiak}, \cite{Ankikumar} 
and \cite{AnkikWarnecke}. 
Giri et al.\cite{Ankikumar} studied the coagulation kernels of the form $K(x,y)=\phi(x)\phi(y)$ for 
some sublinear function $\phi$ under the growth restriction $\phi(x)\leq(1+x)^\mu$ for $0\leq\mu<1$, 
and the selection function $S(x)$ is there also considered under the same growth assumption. 
In \cite{AnkikWarnecke}, Giri et al. proved the existence of solutions to the coagulation equation 
with multifragmentation for a more general fragmentation kernel, in order to cover the fragmentation 
kernel $\Gamma(y,x)=(\alpha + 2)x^\alpha y^{\gamma-(\alpha+1)}$ getting a result for $\alpha>-1$ 
and $\gamma\in]0,\alpha+2[$. The existence proofs in \cite{AnkikWarnecke} and \cite{Ankikumar} are 
based on the well known basic method by Stewart \cite{Stewart}, where the solution is obtained through 
the convergence of the solutions to a sequence of truncated problems. 
In \cite{AnkikWarnecke} the uniqueness of the solutions was not studied. In \cite{BanasiakLamb} Banasiak 
and Lamb proved the existence and uniqueness of solutions to the coagulation-fragmentation equation 
when $K(x,y)\in L_\infty(\mathbb{R}_+\times\mathbb{R}_+)$ while in \cite{LambBanasiak} the authors proved 
existence and uniqueness of classical solutions for the class of coagulation kernels 
$K(x,y)\leq k\big((1+a(x))^\alpha+(1+a(y))^\alpha\big)$ where $a$ is the fragmentation rate, $k>0$, 
and $0\leq\alpha<1$.

To our knowledge there is just one result concerning the coagulation mutlifragmentation equation with 
singular coagulation kernels. Ca\~nizo Rinc\'on \cite{Canizo} proved the existence of 
$L^\infty\big([0,T[,M_1\big)$ solutions in the distribution sense for the coagulation kernels $a(y,y')$
such that
\begin{eqnarray*}
K_a\big(y^\alpha(y')^\beta+(y')^\alpha y^\beta\big)\leq a(y,y')\leq K'_a\big(y^\alpha(y')^\beta+(y')^\alpha y^\beta\big)
\end{eqnarray*}
with $\alpha<\beta<1$, $0<\alpha+\beta<1$, $\beta-\alpha<1$, constants $K_a,K'_a>0$, and where $M_1$ is 
the space of measures $\mu$ on $(0,\infty)$ with first bounded moment, see \cite[Section 3.2]{Canizo} 
or \cite[page 59]{Canizo}. His result is resticted to the kernels with order $\alpha$ and $\beta$ in $y$ 
and $y'$ respectively, and being $\alpha\neq\beta$. The singularity is restricted to the case $\sigma\in]0,1/2[$ 
translated into our terms. The result from \cite{Canizo} leaves out, for example, the cases of the Smoluchowski 
and the equi-partition of kinetic energy kernels. The uniqueness of the solutions was not studied in \cite{Canizo}.

In the present article, our aim is to prove the existence and uniqueness of solutions to the coagulation 
equation with multifragmentation with singular coagulation kernels 
\begin{eqnarray}
K(x,y)\leq k(1+x)^\lambda(1+y)^\lambda(xy)^{-\sigma} \quad\mbox{with}\quad\lambda-\sigma\in[0,1[, \sigma\geq 0,
\label{Fsmolokernel}
\end{eqnarray}
giving in this way an existence and uniqueness result for the case of the important Smoluchowki 
coagulation kernel 
\begin{eqnarray*}
K(x,y)=(x^{1/3}+y^{1/3})(x^{-1/3}+y^{-1/3})
\end{eqnarray*}
for Brownian motion, see Smoluchowski \cite{Smoluchowski}. The equi-partition of kinetic energy (EKE) 
kernel
\begin{eqnarray*}
K(x,y)=(x^{1/3}+y^{1/3})^2\sqrt{\frac{1}{x}+\frac{1}{y}},
\end{eqnarray*}
is also covered by our analysis. We are giving a more general result than Ca\~nizo Rinc\'on \cite{Canizo} 
since we do not restrict our kernel to an specific order. We allow $\alpha$ to be equal $\beta$, the 
singularity can be as big as it is wished, and we obtain \emph{regular} solutions
in the space $C^1\left([0,T],L^1\big(]0,\infty[\big)\right)$ with $T>0$.

Our existence result is based on the proof of Stewart \cite{Stewart}. We extend the methods we developed 
in \cite{Carlos} for singular kernels in the pure coagulation problem.

For our existence and uniqueness result we consider the class of fragmentation kernels
\begin{eqnarray}
\Gamma(y,x)\leq y^\theta b(x,y)
\label{Ffragkernel}
\end{eqnarray}
with
\begin{eqnarray}
\int\limits_0^yb(x,y)x^{-2\sigma}\leq Cy^{-2\sigma}\quad\mbox{for}\quad \theta\in[0,1[, \sigma\in[0,1/2],\;\mbox{and a constant}\; C
\label{Ffragkernel1}
\end{eqnarray}
and such that, there exist $q>1$ and $\tau_1,\tau_2\in[-2\sigma-\theta,1-\theta]$ such that
\begin{eqnarray}
\int\limits_0^yb^q(x,y)\leq B_1y^{q\tau_1},\;\mbox{and}\;\int\limits_0^yx^{-q\sigma}b^q(x,y)\leq B_2y^{q\tau_2}\quad\mbox{for constant $B_1,B_2>0$.}
\label{Ffragkernel2}
\end{eqnarray}
From (\ref{FB}) and (\ref{Ffragkernel}) we have that $S(y)\leq y^\theta$. The case $S(y)=y^\theta$ with $\theta>0$ 
was considered in \cite{McGuinness}, where McGuinness et al.\ studied the pure fragmentation 
equation with singular initial conditions. The selection function $S(y)=y^\theta$ has also been studied 
in \cite{LambBrideGuinness}, \cite{RMZiff}, and \cite{ZiffGrady}. In \cite{McGrady} it has been considered 
for $\theta=0$. 

The class of frangmentation kernels (\ref{Ffragkernel}) holding (\ref{Ffragkernel1}) and (\ref{Ffragkernel2}) 
includes the kernel $\Gamma(y,x)=(\alpha + 2)x^\alpha y^{\gamma-(\alpha+1)}$ for $\alpha>2\sigma+\epsilon-1$ 
and $\gamma\in]0,1[$ with $0<\epsilon<\theta$. 
This kernel was studied by Giri et al. \cite{AnkikWarnecke}, where they proved the existence of weak solutions to 
the coagulation equation with multifragmentation, but with nonsingular coagulation kernels.

In order to study the existence of solutions of (\ref{Fproblema})-(\ref{Fcondinicial}), we define for some 
given $\sigma\geq 0$ the space $Y$ to be the following Banach space with norm $\|\cdot\|_Y$
\begin{eqnarray*}
Y=\left\{u\in L^1(]0,\infty[): \|u\|_Y<\infty\right\}\quad\mbox{where}\quad\|u\|_Y=\int\limits_0^\infty(x+x^{-2\sigma})|u(x)|dx.
\end{eqnarray*}
That $Y$ is a Banach space is easily seen. We also write
\begin{eqnarray*}
\|u\|_x=\int\limits_0^\infty x|u(x)|dx\quad\mbox{and}\quad\|u\|_{x^{-2\sigma}}=\int\limits_0^\infty x^{-2\sigma}|u(x)|dx,
\end{eqnarray*}
and set
\begin{eqnarray*}
Y^+=\left\{u\in Y:u\geq 0 \quad a.e.\right\}.
\end{eqnarray*}

Now we define a weak solution to problem (\ref{Fproblema})-(\ref{Fcondinicial}) in the same way as 
Stewart \cite{Stewart}:
\begin{dfn} Let $T\in]0,\infty]$. A solution $u(x,t)$ of (\ref{Fproblema})-(\ref{Fcondinicial}) is a 
function $u:[0,T[\longrightarrow Y^+$ such that for a.e. $x\in[0,\infty[$ and $t\in[0,T[$ the following properties 
hold
\begin{description}
\item[(i)]$u(x,t)\geq 0$ for all $t\in[0,\infty[$,
\item[(ii)]$u(x,\cdot)$ is continuous on $[0,T[$,
\item[(iii)]for all $t\in[0,T[$ the following integral is bounded
\begin{eqnarray*}
\int\limits_0^t\int\limits_0^\infty K(x,y)u(y,\tau)\,dy\,d\tau<\infty\quad\mbox{and}\quad\int\limits_0^t\int\limits_x^\infty b(x,y)S(y)u(y,\tau)\,dy\,d\tau<\infty
\end{eqnarray*}
\item[(iv)] for all $t\in[0,T[$, $u$ satisfies the following weak formulation of (\ref{Fproblema})
\begin{eqnarray}
u(x,t)\!\!&\!\!=\!\!&\!\!u(x,0)+\int\limits_0^t\left[\frac{1}{2}\int\limits_0^xK(x-y,y)u(x-y,\tau)u(y,\tau)\,dy-\int\limits_0^\infty K(x,y)u(x,\tau)u(y,\tau)\,dy\right. \nonumber\\
\!\!&\!\!\!\!&\!\!\qquad\qquad\qquad\left.+\int\limits_x^\infty b(x,y)S(y)u(y,t)\,dy-S(x)u(x,t)\right]d\tau.\label{elgranasterisco}
\end{eqnarray}
\end{description}
\label{Fdefinicion}
\end{dfn}
In the next sections we make use of the following hypotheses
\begin{hyp}
\begin{description}
  \item[]
  \item[(H1)]  $K(x,y)$ is a continuous non-negative function on $]0,\infty[\times]0,\infty[$, 
	\item[(H2)]  $K(x,y)$ is a symmetric function, i.e. $K(x,y)=K(y,x)$ for all $x,y\in]0,\infty[$, 
	\item[(H3)]  $K(x,y)\leq \kappa(1+x)^\lambda(1+y)^\lambda(xy)^{-\sigma}$ for $\lambda-\sigma\in[0,1[,\sigma\geq 0$, and constant $\kappa$,
	\item[(H4)]  $S(x):]0,\infty[\rightarrow [0,\infty[$ is a continuous non-negative function such that $0\leq S(x)\leq y^\theta$ for $\theta\in[0,1[$,
	\item[(H5)]  $b(x,y)$ is such that $\int\limits_0^yb(x,y)x^{-2\sigma}dx\leq Cy^{-2\sigma}$,
	\item[(H6)]  there exist $q>1$ and $\tau_1,\tau_2\in[-2\sigma-\theta,1-\theta]$ such that
	\begin{eqnarray*}
	\int\limits_0^yb^q(x,y)\leq B_1y^{q\tau_1},\;\mbox{and}\;\int\limits_0^yx^{-q\sigma}b^q(x,y)\leq B_2y^{q\tau_2}\quad\mbox{for constant $B_1,B_2>0$}
	\end{eqnarray*}
\end{description}
\label{Fhypo}
\end{hyp}
In the rest of the paper we consider $\kappa=1$ for the simplicity.

We study the uniqueness of the solutions to (\ref{Fproblema})-(\ref{Fcondinicial}) under the following further 
hypotheses
\begin{description}
  \item[(H3')] $K(x,y)\leq\kappa_1(x^{-\sigma}+x^{\lambda-\sigma})(y^{-\sigma}+y^{\lambda-\sigma})$ \textit{such that} $\sigma\geq 0$, $\lambda-\sigma\in[0,1/2]$, \textit{and} $\kappa_1>0$,
  \item[(H4')] $S(x):]0,\infty[\rightarrow [0,\infty[$ is a continuous non-negative function such that $S(x)=y^\theta$ \textit{for} $\theta\leq\lambda-\sigma$.
\end{description}

The restriction $\lambda-\sigma\in[0,1/2]$ in \textbf{(H3')} limits our uniqueness result to a subset of the 
kernels of the class defined in \textbf{(H3)}, namely to the ones for which $\lambda-\sigma\in[0,1/2]$ holds.  
The restriction $\theta\leq\lambda-\sigma$ in \textbf{(H4')} limits our uniqueness result to a more restricted 
class of fragmentation kernels.

We introduce now some easily derived inequalities that will be used throughout the paper. The proof of 
these inequalities can be found in Giri \cite{AnkikThesis}. For any $x,y>0$
\begin{eqnarray}
2^{p-1}(x^p + y^p)\leq (x + y)^p \leq x^p + y^p\quad &\mbox{if} &\quad 0\leq p\leq 1, \label{Finequality1}\\ 
2^{p-1}(x^p + y^p)\geq (x + y)^p\geq x^p + y^p\quad &\mbox{if} &\quad p\geq 1, \label{Finequality2}\\ 
2^{p-1}(x^p + y^p)\geq (x + y)^p\quad &\mbox{if} &\quad p < 0. \label{Finequality3}
\end{eqnarray}

The paper is organized as follows. In Section $2$ we define the sequence of truncated problems and prove 
in Theorem \ref{Fexistencetruncated} the existence and uniqueness of solutions to them. We extract a weakly 
convergent subsequence in $L^1$ from a sequence of unique solutions for truncated equations to 
(\ref{Fproblema})-(\ref{Fcondinicial}). In Section $3$ we show that the solution of (\ref{Fproblema}) is 
actually the limit function obtained from the weakly convergent subsequence of solutions of the truncated 
problem. In Section $4$ we prove the uniqueness, based on the method of Stewart \cite{Stewartuniq}, of the 
solutions to (\ref{Fproblema})-(\ref{Fcondinicial}) for a modification of the classes (\ref{Fsmolokernel}) 
and (\ref{Ffragkernel}) of coagulation and fragmentation kernels respectively. We obtain uniqueness for 
some kernels which are not covered by the existence result.
%
\section{The Truncated Problem}
$\quad$

We prove the existence of a solution to the problem (\ref{Fproblema})-(\ref{Fcondinicial}) by taking the limit 
of the sequence of solutions of the equations given by replacing the kernel $K(x,y)$ and the selection function 
$S(x)$ by their respective 'cut-off' kernel $K_n(x,y)$ and $S_n(x)$ for any given $n\in\mathbb{N}$
\begin{eqnarray}
\begin{array}{ll}
K_n(x,y)=\left\{
                  \begin{array}{ll}
	                  K(x,y) & \mbox{if}\quad  x+y\leq n\;\;\mbox{and}\;\; x,y\geq \sigma/n \\ 
	                  0      & \mbox{otherwise}.
                  \end{array}
  \right. & S_n(x)=\left\{
                  \begin{array}{ll}
	                  S(x) & \mbox{if}\quad  x\leq n \\ 
	                  0      & \mbox{otherwise}.
                  \end{array}
  \right.
\end{array}
\label{FKnSn}
\end{eqnarray}
For the defined kernels the resulting equations are written as
\begin{eqnarray}
  \dfrac{\partial u^n(x,t)}{\partial t}\!\!&\!\!=\!\!&\!\!\dfrac{1}{2}\int\limits_0^x{K_n(x-y,y)u^n(x-y,t)u^n(y,t)}\,dy - \int\limits_0^{n-x}{K_n(x,y)u^n(x,t)u^n(y,t)}\,dy \nonumber\\
  \!\!&\!\!\!\!&\!\!+\int\limits_x^nS_n(y)b(x,y)u^n(y,t)\,dy - S_n(x)u^n(x,t),
\label{Fcen} 
\end{eqnarray}
with the truncated initial data
\begin{eqnarray}
  u^n_0(x)=\left\{
                   \begin{array}{ll}
                     u_0(x) & \mbox{if}\quad 0\leq x\leq n \\ 
                     0      & \mbox{otherwise},
                   \end{array}
  \right. 
\label{Ficen1}                
\end{eqnarray}
where $u^n$ denotes the solution of the problem (\ref{Fcen})-(\ref{Ficen1}) for $x\in[0,n]$.
\begin{thm}
  Suppose that \textbf{(H1)-(H6)} hold and $u_0\in Y^+$. Then for each $n=2,3,4,\ldots$ the problem 
  (\ref{Fcen})-(\ref{Ficen1}) has a unique solution $u^n$ with $u^n(x,t)\geq 0$ for a.e. $x\in[0,n]$ 
  and $t\in [0,\infty[$. Moreover, for all $t\in [0,\infty[$
  \begin{equation}
    \int_0^n{xu^n(x,t)}\,dx=\int_0^n{xu^n(x,0)}\,dx.
    \label{Fconservation}
  \end{equation}
  \label{Fexistencetruncated}
\end{thm}
The proof of Theorem \ref{Fexistencetruncated} follows proceeding as in \cite[Theorem 3.1]{Stewart}.
%
\subsection{Properties of the solutions of the truncated problem}
\begin{lem}
Let  $u^n$ a solution of the truncated problem (\ref{Fcen})-(\ref{Ficen1}). Then for $\alpha\geq 0$ and 
$n=1,2,\ldots$ we obtain the inequality
\begin{eqnarray*}
\frac{d}{dt}\int\limits_0^nu^n(x,t)x^{-\alpha}dx\leq\int\limits_0^n\int\limits_0^yS_n(y)b(x,y)u^n(y,t)x^{-\alpha}dx\,dy.
\end{eqnarray*}
\label{Fdesigualdad}
\end{lem}
\textbf{Proof.} Multiplying equation (\ref{Fcen}) by $x^{-\alpha}$ and integrating w.r.t $x$ from $0$ 
to $n$, changing the order of integration, then a change of variable $x-y=z$, and again re-changing the 
order of integration while replacing $z$ by $x$ gives
\begin{eqnarray}
\frac{d}{dt}\int\limits_0^nu^n(x,t)x^{-\alpha}dx\!\!&\!\!=\!\!&\!\!\frac{1}{2}\int\limits_0^n\int\limits_0^{n-x}K_n(x,y)u^n(x,t)u^n(y,t)\left[(x+y)^{-\alpha}-x^{-\alpha}-y^{-\alpha}\right]dy\,dx \nonumber\\
                                           \!\!&\!\!\!\!&\!\! +\int\limits_0^n\int\limits_x^nS_n(y)b(x,y)u^n(y,t)x^{-\alpha}dy\,dx-\int\limits_0^nS_n(x)u^n(x,t)x^{-\alpha}dx.\label{22nuevo}
\end{eqnarray}
Using the fact that $x^{-\sigma}$ is a sublinear function and therefore $(x+y)^{-\alpha}-x^{-\alpha}-y^{-\alpha}\leq 0$ we get
\begin{eqnarray*}
\frac{d}{dt}\int\limits_0^nu^n(x,t)x^{-\alpha}dx\leq\int\limits_0^n\int\limits_x^nS_n(y)b(x,y)u^n(y,t)x^{-\alpha}dy\,dx=\int\limits_0^n\int\limits_0^yS_n(y)b(x,y)u^n(y,t)x^{-\alpha}dx\,dy, 
\end{eqnarray*}
which complete the proof of the theorem.\hfill{$\Box$}

In the rest of the paper we consider for each $u^n$ their zero extension on $\mathbb R$, i.e.
\begin{eqnarray*}
\hat{u}^n(x,t)=\left\{
                      \begin{array}{ll}
                        u^n(x,t) & 0\leq x\leq n, \;\; t\in[0,T],\\
                        0        & x<0\;\;\mbox{or}\;\;x>n.
                      \end{array}
               \right.
\end{eqnarray*}
For clarity we drop the notation $\hat{\cdot}$ for the remainder of the paper.
\begin{lem}
Assume that \textbf{(H1)-(H6)} hold. We take $u^n$ to be the non-negative zero extension of the solution 
to the truncated problem found in Theorem \ref{Fexistencetruncated}. Fix $T>0$ and let us define 
$$L(T)=\left(e^{NT}(N+1)+e^{CT}(C+1)+1\right)\|u_0\|_Y.$$ Then the following are true:

\begin{description}
\item[(i)]We have the bound 
\begin{eqnarray*}\int\limits_0^\infty{(1+x+x^{-2\sigma})u^n(x,t)}\,dx \leq L(T)\quad\mbox{for all}\quad t\in[0,T].
\end{eqnarray*} 
\item[(ii)]Given $\epsilon>0$ there exists an $R>1$ such that for all $t\in[0,T]$
\begin{eqnarray*}
\sup_n\left\{\int\limits_R^\infty(1+x^{-\sigma})u^n(x,t)\,dx\right\}\leq\epsilon.
\end{eqnarray*}
\item[(iii)]Given $\epsilon>0$ there exists a $\delta>0$ such that for all $n=2,3,\ldots$ and $t\in\left[0,T\right]$
\begin{eqnarray}
\int\limits_A (1+x^{-\sigma})u^n(x,t)\,dx<\epsilon \quad\quad\mbox{whenever}\quad\quad \mu(A)<\delta,
\label{prop3}
\end{eqnarray}
where $\mu(\cdot)$ denotes the Lebesgue measure.
\end{description}
\label{Flemaproperty}
\end{lem}
\textbf{Proof.}$\;$\textbf{Property (i)} 
Computing the term with the weight $x^{-2\sigma}$ using Lemma \ref{Fdesigualdad} for $\alpha=2\sigma$ and using \textbf{(H5)} we get
\begin{eqnarray*}
\frac{d}{dt}\int\limits_0^nu^n(x,t)x^{-2\sigma}dx\!\!&\!\!\leq\!\!&\!\!\int\limits_0^n\int\limits_0^yS(y)b(x,y)u^n(y,t)x^{-2\sigma}dx\,dy \\
                                                 \!\!&\!\!\leq\!\!&\!\!C\int\limits_0^ny^{\theta-2\sigma}u^n(y,t)\,dy \\
                                                 \!\!&\!\!\leq\!\!&\!\!C\int\limits_0^1y^{-2\sigma}u^n(y,t)\,dy +C\int\limits_1^nyu^n(y,t)\,dy\\
                                                 \!\!&\!\!\leq\!\!&\!\!C\int\limits_0^nx^{-2\sigma}u^n(x,t)\,dx +C\|u_0\|_Y.
\end{eqnarray*}
From this inequality we find as above that
\begin{eqnarray}
\int\limits_0^nu^n(x,t)x^{-2\sigma}dx\leq e^{Ct}(C+1)\|u_0\|_Y, \qquad t\in[0,T].
\label{F2}
\end{eqnarray}
Now, since $1\leq x+x^{-2\sigma}$, by the mass conservation property (\ref{Fconservation}) and by (\ref{F2}) we obtain
\begin{eqnarray*}
\int\limits_0^\infty(1+x+x^{-2\sigma})u^n(x,t)\;dx\!\!&\!\!\leq\!\!&\!\!2\int\limits_0^n(x+x^{-2\sigma})u^n(x,t)\;dx\\
                                            \!\!&\!\!\leq\!\!&\!\!2\left(\|u_0\|_Y+e^{Ct}(C+1)\|u_0\|_Y\right) \\
                                            \!\!&\!\!\leq\!\!&\!\!2\left(e^{CT}(C+1)+1\right)\|u_0\|_Y=: L(T).
\end{eqnarray*}
\textbf{Property (ii)} Choose $\epsilon >0$ and let $R>1$ be such that $R>\frac{2\|u_0\|_Y}{\epsilon}$. Then using (\ref{Fconservation}) we get 
\begin{eqnarray*}
\int\limits_R^\infty(1+x^{-\sigma})u^n(x,t)\,dx = 2\int\limits_R^\infty\frac{x}{x}u^n(x,t)\,dx
\leq \frac{2}{R}\int\limits_0^n xu^n(x,t)\,dx\leq\frac{2}{R}\|u^n_0\|_Y\leq\frac{2}{R}\|u_0\|_Y<\epsilon.
\end{eqnarray*}
\textbf{Property (iii)} 
Let $\chi_A$ denote the characteristic function of a set $A$
and set
\begin{eqnarray}
\kappa(r):=\frac{1}{2}(1+r^\sigma)(1+r)^{2\lambda}.
\label{Fkappa}
\end{eqnarray}
Let us define for all $n=1,2,3,\ldots$ and $t\in[0,T]$ using property \textbf{(i)}
\begin{eqnarray}
f^n(\delta,r,t)=\sup\left\{\int\limits_0^r\chi_{A}(x)(1+x^{-\sigma})u^n(x,t)\,dx:\; A\subset]0,r[\;\mbox{and}\;\mu(A)<\delta\right\}\leq L(T).
\label{Fcond10}
\end{eqnarray}
In the rest of the paper we drop the argument $r$ from $f^n$ for simplicity.

We take $t=0$ in the definition of $f^n$ and observe that $u^n(x,0)\leq u_0(x)$ pointwise almost everywhere. 
Then by the absolute continuity of the Lebesgue integral, we have that
\begin{eqnarray}
f^n(\delta,0)=\sup\left\{\int\limits_0^r\chi_{A}(x)(1+x^{-\sigma})u_0^n(x)\,dx:\; A\subset]0,r[\;\mbox{and}\;\mu(A)<\delta\right\}\rightarrow 0\;\mbox{as $\delta\rightarrow 0$}
\label{Fcond1}
\end{eqnarray}
Now we multiply (\ref{Fcen}) by $(1+x^{-\sigma})\chi_{A}(x)$. This we integrate from $0$ to $t$ w.r.t. $s$ 
and over $[0,r[$ w.r.t. $x$. Using the non-negativity of each $u^n$ we obtain
\begin{eqnarray}
\!\!&\!\!\!\!&\!\!\int\limits_0^r\chi_{A}(x)(1+x^{-\sigma})u^n(x,t)\,dx \nonumber\\ 
\!\!&\!\!\!\!&\!\!\qquad\leq\frac{1}{2}\int\limits_0^t\int\limits_0^r\int\limits_0^x\chi_{A}(x)(1+x^{-\sigma})K_n(x-y,y)u^n(x-y,s)u^n(y,s)\,dy\,dx\,ds\label{diferenciapropiedad}\\ 
\!\!&\!\!\!\!&\!\!\qquad\quad+ \int\limits_0^t\int\limits_0^r\chi_{A}(x)\int\limits_x^nS_n(y)b(x,y)(1+x^{-\sigma})u^n(y,s)\,dy\,dx\,ds + \int\limits_0^r\chi_{A}(x)(1+x^{-\sigma})u^n_0(x)\,dx.\nonumber
\end{eqnarray}
Let us denote $I_{21}$ and $I_{22}$ the first and the second integral terms on the right hand side of 
(\ref{diferenciapropiedad}) respectively. By changing variables as we did in (\ref{22nuevo}) in $I_{21}$ we get
\begin{eqnarray*}
I_{21}(t)=\frac{1}{2}\int\limits_0^t\int\limits_0^r\int\limits_0^{r-y}\chi_{A}(x+y)[1+(x+y)^{-\sigma}]K_n(x,y)u^n(x,s)u^n(y,s)\,dx\,dy\,ds. \nonumber\\ 
\end{eqnarray*}
By using \textbf{(H3)} for $K(x,y)$, then taking $1+(x+y)^{-\sigma}\leq 1+y^{-\sigma}$ and $x^{-\sigma}\leq 1+x^{-\sigma}$ we have
\begin{eqnarray*}
I_{21}(t)\!\!&\!\!\leq\!\!&\!\!\frac{1}{2}\int\limits_0^t\int\limits_0^r\int\limits_0^{r-y}\chi_{A}(x+y)\left[1+(x+y)^{-\sigma}\right](1+x )^\lambda(1+y)^\lambda(xy)^{-\sigma}u^n(x,s)u^n(y,s)\,dx\,dy\,ds \\
\!\!&\!\!\leq\!\!&\!\!\frac{1}{2}\int\limits_0^t\int\limits_0^r\int\limits_0^{r-y}\chi_{A}(x+y)(1+y^{-\sigma})(1+x )^\lambda(1+y)^\lambda(1+x^{-\sigma})y^{-\sigma} u^n(x,s)u^n(y,s)\,dx\,dy\,ds \\
\!\!&\!\!=\!\!&\!\!\frac{1}{2}\int\limits_0^t\int\limits_0^{r}\int\limits_0^{r-y}\chi_{A}(x+y)(1+y^{\sigma})(1+x )^\lambda(1+y)^\lambda(1+x^{-\sigma})y^{-2\sigma} u^n(x,s)u^n(y,s)\,dx\,dy\,ds.
\end{eqnarray*}
By using the definition (\ref{Fkappa}) of $\kappa(r)$ we obtain the following estimates for $I_{21}$
\begin{eqnarray*}
I_{21}(t)\!\!&\!\!\leq\!\!&\!\! \kappa(r)\int\limits_0^t\int\limits_0^ru^n(y,s)y^{-2\sigma}\int\limits_0^\infty\chi_{A-y\cap[0,r-y]}(x)(1+x^{-\sigma})u^n(x,s)\,dx\,dy\,ds \nonumber\\
\end{eqnarray*}
where $A-y:=\left\{\omega>0:\omega=x-y\,\mbox{for some}\,x\in A\right\}$. Since $A-y\cap[0,r-y]\subset [0,r]$ 
and $\mu\left(A-y\cap[0,r-y]\right)\leq\mu(A-y)\leq\mu(A)<\delta$, by using the definition of $f^n$ and
property \textbf{(i)} we have
\begin{eqnarray}
 I_{21}(t)\leq \kappa(r)L(T)\int\limits_0^tf^n(\delta,s)\,ds.
 \label{I21*}
\end{eqnarray}
Working now with the integral term $I_{22}$ we have using \textbf{(H4)} that
\begin{eqnarray*}
I_{22}(t)\!\!&\!\!\leq\!\!&\!\!\int\limits_0^t\int\limits_0^r\chi_{A}(x)\int\limits_x^\infty S_n(y)b(x,y)(1+x^{-\sigma})u^n(y,s)\,dy\,dx\,ds \nonumber\\
&\leq&\int\limits_0^t\int\limits_0^\infty\int\limits_0^y\chi_A(x)S_n(y)b(x,y)(1+x^{-\sigma})u^n(y,s)\,dx\,dy\,ds\nonumber\\
&\leq&\int\limits_0^t\int\limits_0^\infty y^{\theta}u^n(y,s)\left[\int\limits_0^y\chi_A(x)b(x,y)\,dx+\int\limits_0^y\chi_A(x)b(x,y)x^{-\sigma}dx\right]dy\,ds.\nonumber
\end{eqnarray*}
Then by hypotheses \textbf{(H6)} we find
\begin{eqnarray*}
 I_{22}(t)\!\!&\!\!\leq\!\!&\!\!\mu(A)^{\frac{p-1}{p}}\int\limits_0^t\int\limits_0^\infty y^{\theta}u^n(y,s)\left[\left(\int\limits_0^yb^p(x,y)dx\right)^{1/p}+\left(\int\limits_0^yb^p(x,y)x^{-p\sigma}dx\right)^{1/p}\right]dy\,ds\\
 &\leq&\mu(A)^{\frac{p-1}{p}}\int\limits_0^t\int\limits_0^\infty y^{\theta}u^n(y,s)\left(B_1y^{\tau_1}+B_2y^{\tau_2}\right)dy\,ds\leq (B_1+B_2)\mu(A)^{\frac{p-1}{p}}L(T)T.
\end{eqnarray*}
Using the estimates of $I_{21}(t)$ and $I_{22}(t)$ in (\ref{diferenciapropiedad}) we have by taking the supremum
over all A such that $A\subset]0, r[$ with $\mu(A)\leq\delta$
\begin{eqnarray*}
 f^n(\delta,t)\leq\kappa(r)L(T)\int\limits_0^tf^n(\delta,s)ds+(B_1+B_2)L(T)T\delta^{\frac{p-1}{p}}+f^n(\delta,0),\quad t\in[0,T].
\end{eqnarray*}
By using Gronwall's inequality, see e.g.\ Walter \cite[page 361]{Walter}, we get
\begin{eqnarray}
f^n(\delta,t)\leq\left[(B_1+B_2)L(T)T\delta^{\frac{p-1}{p}}+f^n(\delta,0)\right]\exp\big(\kappa(r)L(T)T\big),\quad t\in[0,T].
\label{Ffn}
\end{eqnarray}
Since $f^n(\delta,0)\rightarrow 0$ as $\delta\rightarrow 0$ (\ref{Ffn}) implies that
\begin{eqnarray}
 \lim_{\delta\rightarrow 0}\sup_{n\geq 1, t\in[0,T]}\left\{f^n(\delta,t)\right\}=0.
 \label{propiedad}
\end{eqnarray}
Lemma \ref{Flemaproperty}\textbf{(iii)} is then a consequence of (\ref{propiedad}) and Lemma \ref{Flemaproperty}\textbf{(i)}.
This completes the proof of Lemma \ref{Flemaproperty}. \hfill{$\Box$}

Let us define $v^n(x,t)=x^{-\sigma}u^n(x,t)$. Due to the Lemma \ref{Flemaproperty} above and the 
Dunford-Pettis Theorem \cite[page 274]{Edwards}, we can conclude that for each $t\in[0,T]$ the sequences 
$\big(u^n(t)\big)_{n\in\mathbb N}$ and $\big(v^n(t)\big)_{n\in\mathbb N}$ are weakly relatively compact in $L^1\big(]0,\infty[\big)$.
%
\subsection{Equicontinuity in time}

\begin{lem}
Assume that Hypotheses $1.1$ hold. Take $\big(u^n\big)$ now to be the sequence of extended 
solutions to the truncated problems (\ref{Fcen})-(\ref{Ficen1}) found in Theorem 
\ref{Fexistencetruncated} and $v^n(x,t)=x^{-\sigma}u^n(x,t)$. Then there exists a subsequences 
$\big(u^{n_k}(t)\big)$ and  $\big(v^{n_l}(t)\big)$ of $\big(u^{n}(t)\big)_{n\in\mathbb N}$ and 
$\big(v^{n}(t)\big)_{n\in\mathbb N}$ respectively such that
\begin{eqnarray*}
&& u^{n_k}(t)\rightharpoonup u(t)\quad\mbox{in}\quad L^1\big(]0,\infty[\big)\quad\mbox{as}\quad n_k\rightarrow\infty \\ 
&& v^{n_l}(t)\rightharpoonup v(t)\quad\mbox{in}\quad L^1\big(]0,\infty[\big)\quad\mbox{as}\quad n_l\rightarrow\infty 
\end{eqnarray*}
uniformly for $t\in [0,T]$. This convergence is uniform for all $t\in [0; T]$ giving 
$u,v\in C\left([0,T];\Omega\right)=\left\{\eta:[0,T]\rightarrow \Omega,\right.$
$\left.\eta \;\mbox{continuous and}\; \eta(t)\; \mbox{bounded for all}\; t\in[0,T] \right\}$, 
where $\Omega$ is $L^1\big(]0,\infty[\big)$ equipped with the weak topologyy.\label{Funlema}
\end{lem}
\textbf{Proof}: Choose $\epsilon>0$, and $\phi\in L^\infty\big(]0,\infty[\big)$. Let $s,t\in[0,T]$ and 
assume that $t\geq s$. Choose $a>1$ such that 
\begin{eqnarray}
\frac{2L(T)}{a}\|\phi\|_{L^\infty(]0,\infty[)}\leq\epsilon/2.
\label{F20}
\end{eqnarray}
Let us define the function $\omega(x,t):=u^n(x,t)x^{-\beta}$ for $\beta=0$ or $\beta=\sigma$. Note that 
for $\beta=0$ and $\beta=\sigma$ it becames $u^n(x,t)$ and $v^n(x,t)$ respectively.
Using Lemma \ref{Flemaproperty}, for each $n$, we get
using $a>1$ chosen to satisfy (\ref{F20})
\begin{eqnarray}
\int\limits_a^\infty\left|\omega^n(x,t)-\omega^n(x,s)\right|dx\!\!&\!\!=\!\!&\!\!\int\limits_a^\infty\left|x^{-\beta}u^n(x,t)-x^{-\beta}u^n(x,s)\right|dx \nonumber\\
\!\!&\!\!\leq\!\!&\!\!\frac{1}{a}\int\limits_a^\infty x^{1-\beta}\left|u^n(x,t)+u^n(x,s)\right|dx\nonumber\\
                                                  \!\!&\!\!\leq\!\!&\!\!\frac{1}{a}\int\limits_a^\infty x\left|u^n(x,t)+u^n(x,s)\right|dx \leq 2L(T)/a.
\label{Fotratu}
\end{eqnarray}
By using (\ref{Fcen}), (\ref{F20}), (\ref{Fotratu}), for $t\geq s$  we obtain
\begin{eqnarray}
\!\!&\!\!\!\!&\!\!\left|\int\limits_0^\infty\phi(x)\left[\omega^n(x,t)-\omega^n(x,s)\right]dx\right| \nonumber\\
\!\!&\!\!\!\!&\!\!\leq\int\limits_0^a|\phi(x)|\left|\omega^n(x,t)-\omega^n(x,s)\right|dx + \epsilon/2 \nonumber\\         
\!\!&\!\!\!\!&\!\!\leq\|\phi\|_{L^\infty(]0,\infty[)}\int\limits_s^t\left[\frac{1}{2}\int\limits_0^a\int\limits_0^xK_n(x-y,y)u^n(x-y,\tau)u^n(y,\tau)x^{-\beta}dy\,dx\right. \\
\!\!&\!\!\!\!&\!\!\qquad\qquad\qquad\qquad\quad+\int\limits_0^a\int\limits_0^{n-x}K_n(x,y)u^n(x,\tau)u^n(y,\tau)x^{-\beta}dy\,dx \nonumber\\
\!\!&\!\!\!\!&\!\!\qquad\qquad\qquad\qquad\quad\left.+\int\limits_0^a\int\limits_x^nb(x,y)S_n(y)u^n(y,\tau)x^{-\beta}dy\,dx+\int\limits_0^aS_n(x)u^n(x,\tau)x^{-\beta}dx\right]d\tau + \epsilon/2 \nonumber\\ 
\!\!&\!\!\!\!&\!\!=\|\phi\|_{L^\infty(]0,\infty[)}\int\limits_s^t\left(I_{41}(\tau)+I_{42}(\tau)+I_{43}(\tau)+I_{44}(\tau)\right)d\tau +\epsilon/2. \nonumber
\label{f28}
\end{eqnarray}
A change of variables in the first integral gives

\begin{eqnarray*}
 I_{41}(\tau)=\frac{1}{2}\int\limits_0^a\int\limits_0^{a-x}K_n(x,y)u^n(x,\tau)u^n(y,\tau)(x+y)^{-\beta}dy\,dx.
\end{eqnarray*}
Taking $y=0$ in the term $(x+y)^{-\beta}$ we find that
\begin{eqnarray}
 I_{41}(\tau)\leq \frac{1}{2}\int\limits_0^a\int\limits_0^{a-x}K_n(x,y)u^n(x,\tau)u^n(y,\tau)x^{-\beta}dy\,dx.
\label{Farriba5} 
\end{eqnarray}
By using the definition of $K_n(x,y)$, Lemma \ref{Flemaproperty}\textbf{(i)}, and the fact that $\beta$ just 
take the values $0$ and $\sigma$ we have
\begin{eqnarray}            
I_{41}(\tau)\!\!&\!\!\leq\!\!&\!\!\frac{1}{2}\int\limits_0^a\int\limits_0^{a-x}(1+x)^\lambda(1+y)^\lambda x^{-(\sigma+\beta)}y^{-\sigma}u^n(x,\tau)u^n(x,\tau)\,dy\,dx  \nonumber\\
\!\!&\!\!\leq\!\!&\!\! \frac{1}{2}(1+a)^{2\lambda}\int\limits_0^{a}\int\limits_0^{a-x}x^{-(\sigma+\beta)}y^{-\sigma}u^n(x,\tau)u^n(x,\tau)\,dy\,dx \leq\frac{1}{2}(1+a)^{2\lambda}L(T)^2.
\label{FI41}
\end{eqnarray}
In order to estimate the second term, we define 
\begin{eqnarray}
C_1=\left\{\begin{array}{lcl}
             1 & \mbox{if} & 0\leq\lambda\leq 1 \\
             2^{\lambda-1} & \mbox{if} & \lambda\geq 1.
           \end{array}
    \right.
\label{FC1}    
\end{eqnarray}
Then, by using inequalities (\ref{Finequality1}) and (\ref{Finequality2}) for $p=\lambda$ and 
Lemma \ref{Flemaproperty}\textbf{(i)}, working as in (\ref{FI41}), we find that
\begin{eqnarray}
I_{42}(\tau)\!\!&\!\!\leq\!\!&\!\!C_1^2\int\limits_0^a\int\limits_0^{n-x}(1+x^\lambda)(1+y^\lambda)x^{-(\sigma+\beta)}y^{-\sigma}u^n(x,\tau)u^n(y,\tau)\,dy\,dx \nonumber\\
\!\!&\!\!\leq\!\!&\!\!C_1^2(1+a^\lambda)\int\limits_0^a\int\limits_0^{n-x}(y^{-\sigma}+y^{\lambda-\sigma})x^{-(\sigma+\beta)}u^n(x,\tau)u^n(y,\tau)\,dy\,dx \leq 2C_1^2(1+a^\lambda)L(T)^2.
\label{FI42}
\end{eqnarray}
Now changing the order of integration in $I_{43}(\tau)$ we have
\begin{eqnarray*}
I_{43}(\tau)\!\!&\!\!\leq\!\!&\!\!\int\limits_0^a\int\limits_x^nb(x,y)S(y)u^n(y,\tau)x^{-\beta}dy\,dx \nonumber\\
\!\!&\!\!=\!\!&\!\!\int\limits_0^a\int\limits_0^yb(x,y)y^\theta u^n(y,\tau)x^{-\beta}dx\,dy+\int\limits_a^n\int\limits_0^ab(x,y)y^\theta u^n(y,\tau)x^{-\beta}dx\,dy. \label{F23*}
\end{eqnarray*}
We can see that, by using \textbf{(H5)} and (\ref{FbN}) for $\beta=\sigma$
\begin{eqnarray}
\int\limits_0^yb(x,y)x^{-\beta}dx=\int\limits_0^yb(x,y)x^{-\sigma}dx\!\!&\!\!\leq\!\!&\!\!\int\limits_0^1b(x,y)x^{-2\sigma}dx + \int\limits_1^yb(x,y)dx \nonumber\\
                                  \!\!&\!\!\leq\!\!&\!\!\int\limits_0^yb(x,y)x^{-2\sigma}dx + \int\limits_0^yb(x,y)dx \leq Cy^{-2\sigma}+N
\label{Fprop3}
\end{eqnarray}
and for $\beta=0$
\begin{eqnarray}
\int\limits_0^yb(x,y)x^{-\beta}dx=\int\limits_0^yb(x,y)dx=N.
\label{Fprop33}
\end{eqnarray}
Then, from (\ref{Fprop3}) and (\ref{Fprop33}) we can conclude that 
\begin{eqnarray}
\int\limits_0^yb(x,y)x^{-\beta}dx \leq Cy^{-2\sigma}+N
\label{Fprop333}
\end{eqnarray}
By using (\ref{Fprop333}) and \textbf{(H4)}, $I_{43}(\tau)$ can be estimated by
\begin{eqnarray}
I_{43}(\tau)\!\!&\!\!\leq\!\!&\!\!\int\limits_0^a(Cy^{-2\sigma}+N)y^\theta u^n(y,\tau)\,dy+\int\limits_a^n(Cy^{-2\sigma}+N)y^\theta u^n(y,\tau)\,dy \nonumber\\
\!\!&\!\!=\!\!&\!\!\int\limits_0^n(Cy^{-2\sigma}+N)y^\theta u^n(y,\tau)\,dy\nonumber\\
\!\!&\!\!=\!\!&\!\!C\int\limits_0^ny^{\theta-2\sigma}u^n(y,\tau)\,dy+N\int\limits_0^ny^{\theta}u^n(y,\tau)\,dy\leq(C+N)L(T).
\label{FI43}
\end{eqnarray}
Using \textbf{(H4)} and Lemma \ref{Flemaproperty}\textbf{(i)} we obtain
\begin{eqnarray}
I_{44}(\tau)\leq\int\limits_0^ax^{\theta-\beta} u^n(x,\tau)dx\leq L(T)
\label{FI44}
\end{eqnarray}
which together with (\ref{FI41}), (\ref{FI42}) and (\ref{FI43})  brings (\ref{f28}) to 
\begin{eqnarray}
&&\left|\int\limits_0^\infty\phi(x)\left[\omega^n(x,t)-\omega^n(x,s)\right]dx\right|\nonumber\\
&&\leq \left[\left(\frac{1}{2}(1+a)^{2\lambda}+2C_1^2\left(1+a^\lambda\right)\right)L(T)^2+(C+N+1)L(T)\right](t-s)\|\phi\|_{L^\infty(]0,\infty[)}+\epsilon/2<\epsilon,\label{34}
\end{eqnarray}
whenever $(t-s)<\delta$ for some $\delta>0$ sufficiently small. The argument given above 
similarly holds for $s<t$. Hence (\ref{34}) holds for all $n$ and $|t-s|<\delta$. Then the 
sequence $\big(\omega^n(t)\big)_{n\in\mathbb N}$ is time equicontinuous in $L^1\big(]0,\infty[\big)$.\ Thus, 
$\big(\omega^n(t)\big)$ lies in a relatively compact subset of a gauge space $\Omega$. The gauge 
space $\Omega$ is $L^1\big(]0,\infty[\big)$ equipped with the weak topology. For details about the 
gauge space, see Ash ~\cite[page 226]{Ash}. Then, we may apply a version of the Arzela-Ascoli 
Theorem, see Ash ~\cite[page 228]{Ash}, to conclude that there exists a subsequence 
$\big(\omega^{n_k}\big)_{k\in\mathbb N}$ such that
\begin{eqnarray*}
\omega^{n_k}(t)\rightarrow \omega(t)\quad\mbox{in}\quad \Omega\quad\mbox{as}\quad n_k\rightarrow\infty,
\end{eqnarray*}
uniformly for $t\in [0,T]$ for some $\omega\in C\left([0,T];\Omega\right)$. 

Then taking $\beta=0$ and $\beta=\sigma$ we can
conclude that there exist subsequence $\big(u^{n_k}\big)_{k\in\mathbb N}$ and $\big(v^{n_k}\big)_{k\in\mathbb N}$ such that
\begin{eqnarray*}
u^{n_k}(t)\rightarrow u(t)\quad\mbox{in}\quad \Omega\quad\mbox{as}\quad n_k\rightarrow\infty\\
v^{n_k}(t)\rightarrow v(t)\quad\mbox{in}\quad \Omega\quad\mbox{as}\quad n_k\rightarrow\infty,
\end{eqnarray*}
uniformly for $t\in [0,T]$ for some $u,v\in C\left([0,T];\Omega\right)$. \hfill{$\Box$}

\section{Existence Theorem}
\subsection{Convergence of the integrals}

In order to show that the limit function which we obtained above is indeed a solution to 
(\ref{Fproblema})-(\ref{Fcondinicial}), we define the operators $M^n_i$, $M_i$, $i=1,2,3,4$
\begin{eqnarray*}
\begin{array}{ll}
 M_1^n(u^n)(x)=\frac{1}{2}\int\limits_{0}^xK_n(x-y,y)u^n(x-y)u^n(y)\,dy & M_1(u)(x)=\frac{1}{2}\int\limits_0^xK(x-y,y)u(x-y)u(y)\,dy \\
 M_2^n(u^n)(x)=\int\limits_{0}^{n-x}K_n(x,y)u^n(x)u^n(y)\,dy & M_2(u)(x)=\int\limits_0^\infty K(x,y)u(x)u(y)\,dy \\
 M_3^n(u^n)(x)=\int\limits_x^nb(x,y)S_n(y)u^n(y)\,dy & M_3(u)(x)=\int\limits_x^\infty b(x,y)S(y)u(y)\,dy \\
 M_4^n(u^n)(x)=S_n(x)u^n(x) & M_4(u)(x)=S(x)u(x),
\end{array}
\end{eqnarray*}
where $u\in L^1\big(]0,\infty[\big)$, $x\in]0,\infty[$ and $n=1, 2, \ldots$. Set 
$M^n=M^n_1-M^n_2+M^n_3-M^n_4$ and $M=M_1-M_2+M_3-M_4$.
\begin{lem}
Suppose that $\big(u^n\big)_{n\in\mathbb N}\subset Y^+$, $u\in Y^+$ where 
$\|u^n\|_Y\leq L$, $\|u\|_Y\leq Q$, $u^n\rightharpoonup u$ and $v^n\rightharpoonup v$ in 
$L^1(]0,\infty[)$ as $n\rightarrow\infty$, for $v=x^{-\sigma}u$ and $v^n=x^{-\sigma}u^n$. 
Then for each $a>0$
\begin{eqnarray*}
M^n(u^n)\rightharpoonup M(u)\quad\mbox{in}\quad L^1\big(]0,a[\big)\quad\mbox{as}\quad n\rightarrow\infty. 
\end{eqnarray*}
\label{Flemaconvergence}
\end{lem}
\textbf{Proof}: Choose $a>0$ and let $\phi\in L^\infty\big(]0,\infty[\big)$. We show that 
$M^n_i(u^n)\rightharpoonup M_i(u)$ in $L^1\big(]0,a[\big)$ as $n\rightarrow\infty$ for $i=1,2,3,4$. \\[14pt]
The proof of case $i=1$ is analogous to the proof of the $W_1$ case in Stewart \cite[Lemma 4.1]{Stewart} by taking
\begin{eqnarray*}
h(v)(x)=\frac{1}{2}\int\limits_0^{a-x}\phi(x+y)K(x,y)(xy)^\sigma v(y)\,dy\quad\mbox{where}\quad v=x^{-\sigma}u. 
\end{eqnarray*}
For every $\epsilon>0$ and  $C_1$ defined by (\ref{FC1}) we can choose 
$\eta$ large enough, due to the negative exponents, such that for $L,Q$ from our assumptions 
\begin{eqnarray}
C_1\|\phi\|_{L^\infty([0,a])}\left[(2\eta^{-(1+\sigma)}+\eta^{\lambda-\sigma-1})(L^2+Q^2)\right]<\frac{\epsilon}{3}.
\label{Mcase2}
\end{eqnarray}
Redefining the operator $h$ for $u\in Y^+$ and $x\in[0,a]$ by
\begin{eqnarray*}
h(v)(x)=\int\limits_0^\eta\phi(x)K(x,y)(xy)^\sigma v(y)\,dy.
\end{eqnarray*}
We can now follow the lines of the of the proof of the $W_2$ case in Stewart \cite[Lemma 4.1]{Stewart} 
to get the proof of case $i=2$.

Case $i=3,4$ can be proved analogously as in Giri et al.\ \cite[Lemma 2.2]{AnkikWarnecke}.

Then the proof of Lemma \ref{Flemaconvergence} is complete. \hfill{$\Box$}
%
\subsection{The existence result}

\begin{thm}
Suppose that \textbf{(H1)}-\textbf{(H6)} hold and assume that $u_0\in Y^+$. 
Then (\ref{elgranasterisco}) has a solution $u\in C\left([0,T],L^1\big(]0,\infty[\big)\right)$. 
Moreover, we also obtain $u\in C^1\left([0,T],L^1\big(]0,\infty[\big)\right)$ and therefore $u$
is a \emph{regular} solution satisfying (\ref{Fproblema}).
\label{Fexisten}
\end{thm}
\textbf{Proof.} Choose $T,m > 0$, and let $\big(u^n\big)_{n\in\mathbb N}$ be the weakly 
convergent subsequence of approximating solutions obtained in Lemma \ref{Funlema}. For 
$t\in[0,T]$ we obtain by weak convergence and Lemma \ref{Flemaproperty}\textbf{(i)}
\begin{eqnarray*}
\int\limits_0^mxu(x,t)\,dx=\lim_{n\rightarrow\infty}\int\limits_0^mxu^n(x,t)\,dx\leq L(T)<\infty,
\end{eqnarray*}
and
\begin{eqnarray*}
\int\limits_{1/m}^mx^{-\sigma}u(x,t)\,dx=\lim_{n\rightarrow\infty}\int\limits_{1/m}^mx^{-\sigma}u^n(x,t)\,dx \leq L(T)<\infty.  
\end{eqnarray*}
Then taking $m\rightarrow\infty$ implies that $u\in Y^+$ with $\|u\|_Y\leq 2L(T)$. Let 
$\phi\in L^\infty\big(]0,a[\big)$. From Lemma \ref{Funlema} we have for each $s\in[0,t]$
\begin{eqnarray}
u^n(s)\rightharpoonup u(s)\quad\mbox{in}\quad L^1\big(]0,a[\big)\quad\mbox{as}\quad n\rightarrow\infty.
\label{F3232}
\end{eqnarray}
From Lemma \ref{Funlema} and Lemma \ref{Flemaconvergence} for each $s\in[0,t]$ we have
\begin{eqnarray}
\int\limits_0^a\phi(x)\left[M^n(u^n(s))(x)-M(u(s))(x)\right]\,dx\rightarrow 0\quad\mbox{as}\quad n\rightarrow\infty.
\label{F341}
\end{eqnarray}
Also, for $s\in[0,t]$, using Lemma \ref{Flemaproperty}\textbf{(i)} and $\|u\|_Y\leq 2L(T)$ we find that
\begin{eqnarray*}
\int\limits_0^a\left|\phi(x)\right|\left|M^n(u^n(s))(x)-M(u(s))(x)\right|dx\leq 3\|\phi\|_{L^\infty(]0,a[)}\left[10L(T)+\left(N+1\right)\right]L(T).
\end{eqnarray*}
Then, by (\ref{F341}), the dominated convergence theorem, and Fubini's Theorem we get
\begin{eqnarray}
\int\limits_0^t M^n(u^n(s))(x)\,ds\rightharpoonup\int\limits_0^tM(u(s))(x)\,ds\quad\mbox{in}\quad L^1\big(]0,a[\big)\quad\mbox{as}\quad n\rightarrow\infty.
\label{F371}
\end{eqnarray}
From the definition of $M^n$ for $t\in[0,T]$
\begin{eqnarray*}
u^n(t)=\int\limits_0^t M^n(u^n(s))\,ds + u^n(0),
\end{eqnarray*}
and thus it follows by (\ref{F371}), (\ref{F3232}) and the uniqueness of weak limits that
\begin{eqnarray}
u(t)=\int\limits_0^t M(u(s))\,ds + u(0). 
\label{388}
\end{eqnarray}
It follows from the fact that $a$ is arbitrary that $u$ is a solution to 
(\ref{Fproblema}) on $C\big([0,T],\Omega\big)$. 

Now we show that $u\in C\left([0,T];L^1\big(]0,\infty[\big)\right)$. 
Considering $t_n>t$ and by using (\ref{388}) we have that
\begin{eqnarray*}
\int\limits_0^\infty\left|u(x,t_n)-u(x,t)\right|dx\!\!&\!\!=\!\!&\!\!\int\limits_0^\infty\left|\frac{1}{2}\int\limits_t^{t_n}\int\limits_0^xK(x-y,y)u(x-y,\tau)u(y,\tau)\,dy\,d\tau\right.\\
&&\qquad\qquad-\int\limits_t^{t_n}\int\limits_0^\infty K(x,y)u(x,\tau)u(y,\tau)\,dy\,d\tau\\
&&\qquad\qquad\left.+\int\limits_t^{t_n}\int\limits_x^\infty b(x,y)S(y)u(y,\tau)\,dy\,d\tau-\int\limits_t^{t_n}S(x)u(x,\tau)\,d\tau\right|dx\\
&\leq&\int\limits_t^{t_n}\left[\frac{3}{2}\int\limits_0^\infty\int\limits_0^\infty K(x,y)u(x,\tau)u(y,\tau)\,dy\,dx\right.\\
&&\qquad\qquad\left.+\int\limits_0^\infty\int\limits_0^yb(x,y)S(y)u(y,\tau)\,dx\,dy+\int\limits_0^\infty S(x)u(x,\tau)\,dx\right]d\tau.
\end{eqnarray*}
By using the definition (\ref{FC1}) of $C_1$, Lemma \ref{Flemaproperty} \textbf{(i)}, \textbf{(H3)}, 
\textbf{(H4)} and (\ref{FbN}) we find that
\begin{eqnarray}
\int\limits_0^\infty\left|u(x,t_n)-u(x,t)\right|dx\!\!&\!\!\leq\!\!&\!\!\int\limits_t^{t_n}\left[\frac{3}{2}\int\limits_0^\infty\int\limits_0^\infty (1+x)^\lambda(1+y)^\lambda(xy)^{-\sigma}u(x,\tau)u(y,\tau)\,dy\,dx\right.\nonumber\\
&&\qquad\qquad+\left.N\int\limits_0^\infty yu(y,\tau)\,dy+\int\limits_0^\infty xu(x,\tau)\,dx\right]d\tau\nonumber\\
&\leq&\int\limits_t^{t_n}\left[\frac{3}{2}C_1^2\int\limits_0^\infty\int\limits_0^\infty\Big((xy)^{-\sigma}+x^{\lambda-\sigma}y^{-\sigma}+y^{\lambda-\sigma}x^{-\sigma}\Big)u(x,\tau)u(y,\tau)\,dy\,dx\right.\nonumber\\
&&\qquad\qquad+\left.N\int\limits_0^\infty yu(y,\tau)\,dy+\int\limits_0^\infty xu(x,\tau)\,dx\right]d\tau\nonumber\\
&\leq&\left[\frac{9}{2}C_1^2L^2(T)+(N+1)L(T)\right](t_n-t).
\label{389}
\end{eqnarray}
Then from (\ref{389}) we obtain that
\begin{eqnarray}
\int\limits_0^\infty\left|u(x,t_n)-u(x,t)\right|dx\rightarrow 0\quad\mbox{as}\quad t_n\rightarrow t.
\label{390}
\end{eqnarray}
The same argument holds when $t_n<t$. Hence (\ref{390}) holds for $\left|t_n-t\right|\rightarrow 0$ and we can conclude
that $u\in C\left([0,T];L^1\big([0,\infty[\big)\right)$. 

Now, we have that our solution satisfies the integral equation
\begin{eqnarray}
u(x,t)\!\!&\!\!=\!\!&\!\!u(x,0)+\int\limits_0^t\left[\frac{1}{2}\int\limits_0^xK(x-y,y)u(x-y,\tau)u(y,\tau)\,dy-\int\limits_0^\infty K(x,y)u(x,\tau)u(y,\tau)\,dy\right. \nonumber\\
\!\!&\!\!\!\!&\!\!\qquad\qquad\qquad\left.+\int\limits_x^\infty b(x,y)S(y)u(y,t)\,dy-S(x)u(x,t)\right]d\tau.\label{36}
\end{eqnarray}
From this we can see that for $u$, which is a continuous function in time $t$, that the integrand
\begin{eqnarray}
f(x,t)\!\!&\!\!=\!\!&\!\!\frac{1}{2}\int\limits_0^xK(x-y,y)u(x-y,\tau)u(y,\tau)\,dy-\int\limits_0^\infty K(x,y)u(x,\tau)u(y,\tau)\,dy \nonumber\\
\!\!&\!\!\!\!&\!\!\qquad+\int\limits_x^\infty b(x,y)S(y)u(y,t)\,dy-S(x)u(x,t)\label{37}
\end{eqnarray}
is also a continuous function in time. We now show that $f(\cdot,t)\in L^1\big([0,\infty[\big)$ for any $t\in[0,T]$.

Integrating (\ref{37}) from $0$ to $\infty$ w.r.t. $x$ we have to show that the following integral is bounded
\begin{eqnarray}
\int\limits_0^\infty f(x,t)dx\!\!&\!\!=\!\!&\!\!\frac{1}{2}\int\limits_0^\infty\int\limits_0^xK(x-y,y)u(x-y,\tau)u(y,\tau)\,dy\,dx -\int\limits_0^\infty\int\limits_0^\infty K(x,y)u(x,\tau)u(y,\tau)\,dy\,dx \nonumber\\
\!\!&\!\!\!\!&\!\!\qquad+\int\limits_0^\infty\int\limits_x^\infty b(x,y)S(y)u(y,t)\,dy\,dx-\int\limits_0^\infty S(x)u(x,t)\,dx.\label{38}
\end{eqnarray}
Working with the second, third and fourth terms of the right hand side of (\ref{38}) as in (\ref{389}) we find that
\begin{eqnarray}
&&\int\limits_0^\infty\int\limits_0^\infty K(x,y)u(x,\tau)u(y,\tau)\,dy\,dx<\infty,\label{39-1} \\
&&\int\limits_0^\infty\int\limits_x^\infty b(x,y)S(y)u(y,t)\,dy\,dx=\int\limits_0^\infty\int\limits_0^y b(x,y)S(y)u(y,t)\,dx\,dy<\infty\quad\mbox{and},\label{39-2}\\
&&\int\limits_0^\infty S(x)u(x,t)\,dx<\infty\label{39-3}
\end{eqnarray}
Now, by Tonelli's Theorem \cite[page 293]{Nielsen} we have that
\begin{eqnarray*}
\int\limits_0^\infty\int\limits_0^xK(x-y,y)u(x-y,\tau)u(y,\tau)\,dy\,dx=\int\limits_0^\infty\int\limits_y^\infty K(x-y,y)u(x-y,\tau)u(y,\tau)\,dy\,dx
\end{eqnarray*}
holds if
\begin{eqnarray*}
\int\limits_0^\infty\int\limits_0^xK(x-y,y)u(x-y,\tau)u(y,\tau)\,dy\,dx<\infty
\end{eqnarray*}
or
\begin{eqnarray*}
\int\limits_0^\infty\int\limits_y^\infty K(x-y,y)u(x-y,\tau)u(y,\tau)\,dy\,dx<\infty.
\end{eqnarray*}
Making a change of varible $x-y = x'$, $y = y'$ in the second integral term we find, by using the
symmetry of $K(x, y)$, that using (\ref{39-1})
\begin{eqnarray*}
\int\limits_0^\infty\int\limits_y^\infty K(x-y,y)u(x-y,\tau)u(y,\tau)\,dx\,dy\!\!&\!\!=\!\!&\!\!\int\limits_0^\infty\int\limits_0^\infty K(x',y')u(x',\tau)u(y',\tau)\,dx'\,dy'\nonumber\\
&=&\int\limits_0^\infty\int\limits_0^\infty K(x',y')u(x',\tau)u(y',\tau)\,dy'\,dx'<\infty.
\end{eqnarray*}
From this it follows that
\begin{eqnarray}
\int\limits_0^\infty\int\limits_0^x K(x-y,y)u(x-y,\tau)u(y,\tau)\,dx\,dy\!\!&\!\!=\!\!&\!\!\int\limits_0^\infty\int\limits_0^\infty K(x',y')u(x',\tau)u(y',\tau)\,dy'\,dx'<\infty.
\label{40}
\end{eqnarray}
Then, from (\ref{39-1})-(\ref{39-3}) and (\ref{40}) together with (\ref{38}) it follows 
that $f(\cdot,t)\in L^1\big([0,\infty[\big)$. Moreover, we
have that $f\in C\left([0,T];L^1\big([0,\infty[\big)\right)$. Then, using this fact, 
(\ref{36}), (\ref{37}), and $u(x,0)\in Y^+$ we find that
\begin{eqnarray*}
u(x,t)\!\!&\!\!=\!\!&\!\!u(x,0)+\int\limits_0^t\left[\frac{1}{2}\int\limits_0^xK(x-y,y)u(x-y,\tau)u(y,\tau)\,dy-\int\limits_0^\infty K(x,y)u(x,\tau)u(y,\tau)\,dy\right. \nonumber\\
\!\!&\!\!\!\!&\!\!\qquad\qquad\qquad\left.+\int\limits_x^\infty b(x,y)S(y)u(y,t)\,dy-S(x)u(x,t)\right]d\tau.
\end{eqnarray*}
gives $u\in C^1\left([0,T];L^1\big(]0,\infty[\big)\right)$ since the right hand side lies in 
this space. And this completes the proof of 
Theorem \ref{Fexisten}.\hfill{$\Box$}

\section{Uniqueness of Solutions}
\begin{thm}
If \textbf{(H1)}, \textbf{(H2)}, \textbf{(H3')}, \textbf{(H4')}, and \textbf{(H5)} hold then the problem 
(\ref{Fproblema})-(\ref{Fcondinicial}) has a unique solution $u\in C\left([0,T],L^1\big(]0,\infty[\big)\right)$.
\end{thm}
\textbf{Proof}: Let us consider $u_1$ and $u_2$ to be solutions to (\ref{Fproblema})-(\ref{Fcondinicial}) 
on $[0,T]$ for $T>0$ arbitrarily chosen, with $u_1(x,0)=u_2(x,0)$ and set $U=u_1-u_2$. We define for $n=1,2,3,\ldots$
\begin{eqnarray*}
m^n(t)=\int\limits_0^n (x^{-\sigma}+x^{\lambda-\sigma})\left|U(x,t)\right|dx.
\end{eqnarray*}
Then, proceeding analogously as in \cite{Stewartuniq} we have that 
\begin{eqnarray*}
u_1(x,t)=u_2(x,t)\quad\mbox{for a.e.}\quad x\in]0,\infty[. 
\end{eqnarray*}
For details see Cueto Camejo \cite{CuetoCamejo}. \hfill{$\Box$}
%
\phantomsection
\section*{Acknowledgements}

This work was supported by the International Max-Planck Research School, `Analysis, Design 
and Optimization in Chemical and Biochemical Process Engineering', 
Otto-von-Guericke-Universit\"at Magdeburg. The authors thank gratefully for the funding of C. 
Cueto Camejo through this PhD program by the state of Saxony-Anhalt.

\end{document}